\documentclass[twocolumn,prl,amsmath,amssymb,superscriptaddress]{revtex4-1}
\usepackage{graphicx,bm,dsfont,amsmath,amssymb}
\usepackage{soul}
\usepackage{color}
\usepackage{amsthm}
\usepackage{mathrsfs}
\usepackage{subfigure}
\usepackage[colorlinks=true,citecolor=blue,urlcolor=blue]{hyperref}

\UseRawInputEncoding
\begin{document}

\title{Majorana-like oscillation for edge states in one-dimensional topological chain with dissipative couplings}
\author{Yang Zhang}
\affiliation{School of integrated circuits, Tsinghua University, Beijing 100084, China}
\affiliation{Frontier Science Center for Quantum Information, Beijing, China}
\author{Wei Nie}
\affiliation{School of integrated circuits, Tsinghua University, Beijing 100084, China}
\author{Yu-xi Liu}\email{yuxiliu@mail.tsinghua.edu.cn}
\affiliation{School of integrated circuits, Tsinghua University, Beijing 100084, China}
\affiliation{Frontier Science Center for Quantum Information, Beijing, China}

\begin{abstract}
The oscillation of Majorana modes with near zero energy plays a very important role for ascertaining Majorana fermions.  The edge states, which also have almost-zero-energy in one-dimensional Su-Schrieffer-Heeger chain (SSHc),  have been extensively studied for their topologically protected properties when the on-sites have dissipations induced by independent environments. We here show that common environments shared by each pair of the nearest neighbour sites in the SSHc can result in dissipative couplings between sites, and thus change topologically trivial phase to nontrivial one. The Majorana-like oscillation for the finite-size hybridizations of two non-Hermitian edge states with complex localization lengths can be induced by the dissipative coupling. The controllable topology parameter of the SSHc plays the role of the magnetic field in the nanowire for controlling Majorana oscillation. The measurement for the oscillation is proposed. Our study provides a new way to manipulate edge states and is experimentally feasible within current technology of superconducting quantum circuits.
\end{abstract}

\maketitle

\textit{Introduction}.---Majorana modes, corresponding to zero energy states, are potential candidates for topological quantum computing~\cite{RevModPhys.82.3045,RevModPhys.83.1057,Bernevig2013robustness,Kitaev2003topological,RevModPhys.80.1083,QuantumInf.1.15001},
which provides protection against noise locally acting on individual or small sets of qubits at a hardware level.
Among various works~\cite{PhysRevLett.100.096407,PhysRevLett.105.077001,PhysRevLett.105.177002,Science.336.1003,NanoLett.12.6414,Nature.464.187,NatMat.14.400} for identifying Majorana modes and Majorana fermions~\cite{Majorana1937,NatPhys.5.614,NatRevMater.3.52}, semiconductor-superconductor nanowires receive considerable attention~\cite{NatPhys.6.336,NatPhys.13.563}. There, two-fold degenerate Majorana modes are supposed to be separately localized at the two ends of nanowire. However, the realistic nanowire has finite length, which results in the spatial hybridization of two modes. It was found that the energies of the hybridized modes oscillate~\cite{PhysRevB.82.094504,PhysRevB.86.220506,PhysRevB.97.155425,PhysRevB.87.094518,PhysRevLett.122.147701,PhysRevB.87.024515,Nature.531.206,PhysRevLett.118.137701,Science.357.6348,PhysRevLett.118.137701} in an exponentially decaying way as a function of externally applied magnetic field. Such oscillation could be a signature for the presence of the Majorana modes.

Zero energy states are also present in other topological systems, e.g., a one-dimensional SSHc with finite size, which has two zero energy eigenstates~\cite{LectureNotesVol.919}. These two eigenstates strongly localize at the two edges of the SSHc, and thus are called as edge states or modes, which are topologically protected. Similar to the nanowire, the finite size of the SSHc also results in the overlap between the two edge states, and thus the energies corresponding to the hybridized states are monotonously far away from zero when the topological parameter approaches the phase transition point.

Recently, environmental effect on topological physics~\cite{ZhaoPRA2017,CuiPRA2020} has been studied via various complex on-site potentials including dissipations or/and gains~\cite{ZhuPRA2014,PhysRevA.101.013635,PhysRevB.97.115436,PhysRevB.101.235150,PhysRevB.100.205119,PhysRevA.99.042111,PhysRevLett.123.066404}, from independent environments~\cite{Hamazaki2019manybody,PhysRevA.98.013628,PhysRevLett.102.065703,PhysRevLett.115.040402,Xiao2017,Leykam2021}. 
The common environment, which is explored for decoherence-free codes~\cite{LMPRL1997,ZanardiPRL1997,DAPRL1998,WuPRL2002}, can result in both the local dissipative potential and the correlated dissipative coupling~\cite{Nature594.369}. The dissipative coupling~\cite{PhysRevA.98.063815,PhysRevLett.123.127202,OE.27.013858} has been applied to study level attraction~\cite{PRL121Harder,PRA13Zhao,PRA102Peng,PRA104Jiang}, light amplification and absorption~\cite{PhysRevX.5.021025,PhysRevLett.122.143901,Wanjura2020,PhysRevApplied.15.044041}. However, the dissipative coupling effect on the topological phase is less studied.

Considering significant progress on topological simulations using superconducting qubits~\cite{RevModPhys.88.021004,PhysRevLett.62.2747,PhysRevLett.104.056404,Science.358.1175,PhysRevLett.120.050507,TanPRL120,TanPRL122,HanPRL2019}, we here study a physically realizable SSHc formed by superconducting qubits~\cite{RepProgPhys.74.104401,PhysRep.718.1}, in which each qubit has an independent environment and also shares common environments with its nearest neighbour qubits. We find that the common-environment-induced dissipative couplings can nontrivially change the topological phase and give rise to non-Hermitian edge states with complex localization lengths. The interaction between edge states exhibits Majorana-like oscillation. The topology-control parameter of the SSHc plays the role of the magnetic field for nanowires in controlling Majorana oscillation. The effect of different on-site complex potentials on the oscillation is analyzed.

\begin{figure}
\includegraphics[scale=0.18]{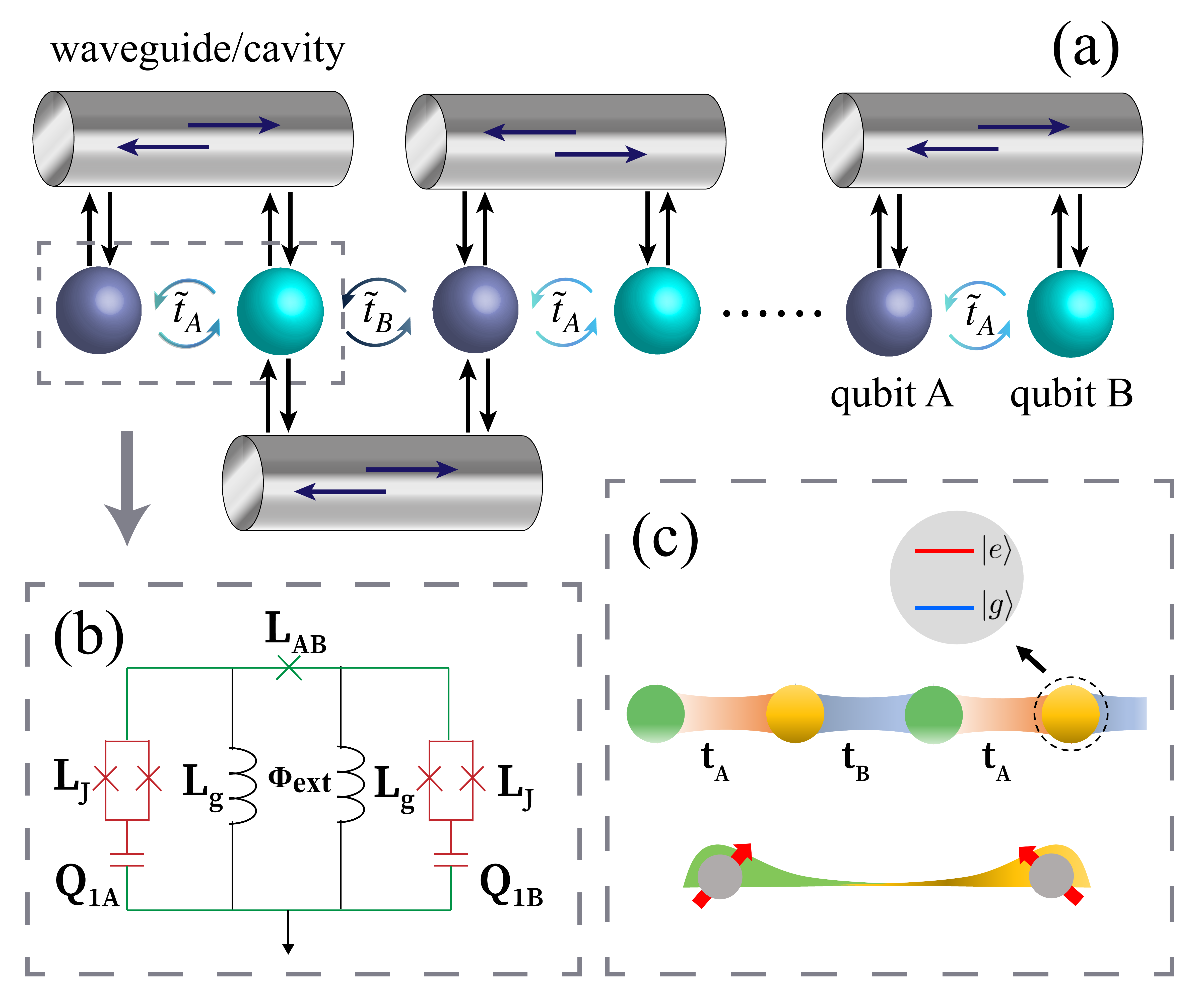}
\caption{(a) Schematics of the SSHc formed by superconducting qubits, in which the nearest neighbour A-type (blue) and B-type (cyan) qubits are coupled to either single-mode microwave cavities or waveguides, mimicking common environments. The circuit realization of the tunable intracell coupling $\tilde{t}_{A}$ between the qubits $A$ and $B$, realized by a magnetic flux biased Josephson junction, is shown in (b). The intercell coupling $\tilde{t}_{B}$ can be realized in a similar way as for $\tilde{t}_{A}$. (c) Schematics of the equivalent SSHc formed by A-type (green) and B-type (yellow) qubits with dissipative couplings $t_{A}=\tilde{t}_{A}-i\gamma_1$ and $t_{B}=\tilde{t}_{B}-i\gamma_2$. Middle figure in (c), colour change for the balls in the chain implies that the qubits in (a) undergo dissipations. Lower figure in (c) denotes the indirect coupling between the first and last qubits, induced by the qubits in the middle of the chain. Upper figure in (c) implies that each ball denotes a two-level system, i.e., qubit.}\label{figure1}
\end{figure}

\textit{Model}.---As shown in Fig.~\ref{figure1}(a), we assume that a SSHc with $N$ unit cells is realized by superconducting qubits~\cite{PhysRep.718.1}, with the identical frequency $\omega$. Each pairs of the nearest neighbour qubits in the SSHc are assumed to have common environments, mimicked by either bad single-mode microwave cavities or waveguides. The intracell and intercell coupling constants between qubits are $\tilde{t}_A=t_0(1-\cos\varphi)$ and $\tilde{t}_B=t_0(1+\cos\varphi)$ with the controllable parameter $\varphi$,  implemented in different ways~\cite{PhysRevLett.90.127901,PhysRevLett.96.067003,PhysRevB.74.172505,PhysRevLett.98.057004,Nature.449.443,PhysRevB.73.094506,Science.316.723,PhysRevB.73.094506,Science.316.723,PhysRevApplied.10.054062,PhysRevResearch.2.012076}, e.g., a magnetic flux biased Josephson junction, as shown in Fig.~\ref{figure1}(b). By eliminating modes of the common environment, as shown in Fig.~\ref{figure1}(c) and Ref.~\cite{SupplementalMaterial}, we obtain the effective non-Hermitian Hamiltonian $H_{\rm eff}=H_{0}+H_{\rm in}$ with $H_{0}=\sum_{n=1}^{N} \omega(\hat{a}_{n}^\dagger \hat{a}_{n} + \hat{b}_{n}^\dagger \hat{b}_{n})$ and

\begin{equation}
\begin{split}
H_{\rm in}=\sum_{n=1}^{N} t_A\left(\hat{a}_{n}^\dagger \hat{b}_{n} + {\rm H.c.}\right) + \sum_{n=1}^{N-1} t_B\left(\hat{a}_{n+1}^\dagger \hat{b}_{n} + {\rm H.c.}\right).   \label{array_Hamiltonian}
\end{split}
\end{equation}
Here, we assume $\hbar=1$ and the complex couplings are $t_A=\tilde{t}_{A}-i\gamma_1$ and $t_{B}=\tilde{t}_{B}-i\gamma_2$. Operators $\hat{a}_n^\dagger=|A_n\rangle \langle \alpha_n|$ and $\hat{b}_n^\dagger=|B_n\rangle\langle\beta_n|$ denote the raising operators of the qubits A and B at the $n$th unit cell with the ground (excited) states $|\alpha_n\rangle(|A_n\rangle)$ and $|\beta_n\rangle(|B_n\rangle)$, respectively. All qubits A (B) are assumed to have the same dissipative constant $\gamma_1$ ($\gamma_2$). Here, to highlight the role of the dissipative couplings, the on-site dissipations of qubits are neglected. This issue will be analyzed later and the detail is given in Ref.~\cite{SupplementalMaterial}. Hereafter, for the convenience, the Hamiltonian of the SSHc with (without) the dissipative coupling is called as the non-Hermitian (Hermitian) Hamiltonian.

\begin{figure}
\includegraphics[scale=0.22]{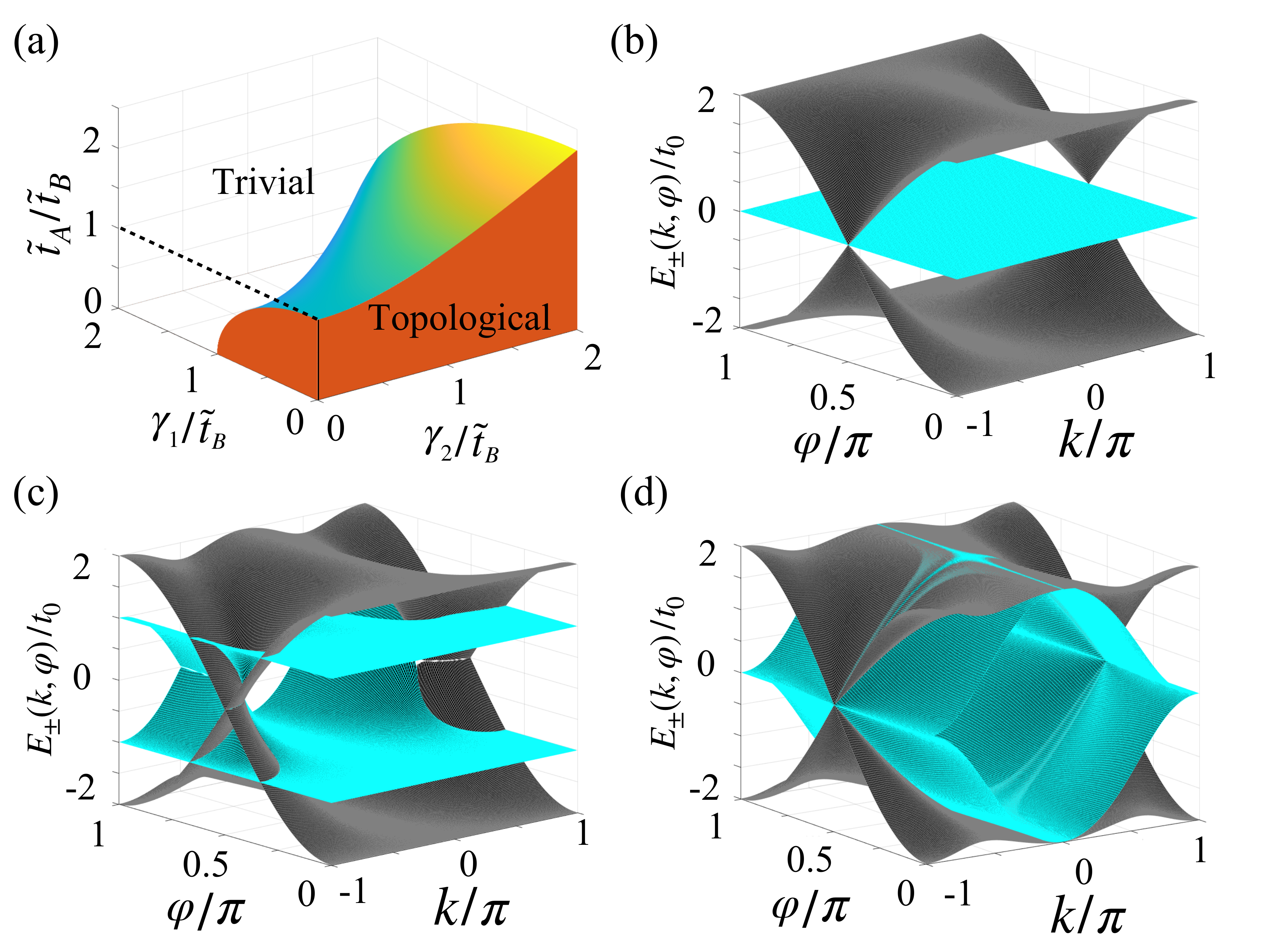}
\caption{(a) Phase diagram for the SSHc with the dissipative couplings. The black solid line segment denotes the topological regime of the Hermitian Hamiltonian.  Real (dark grey) and imaginary parts (cyan) of the dispersion relation $E_{\pm}(k)$ with $\gamma_{1}=\gamma_{2}=0$ in (b), $\gamma_{1}=0, \,\gamma_{2}/t_{0}=1$ in (c), and  $\gamma_{1}/t_{0}=\gamma_{2}/t_{0}=1$ in (d).}\label{figure2}
\end{figure}

\textit{Phase transition point and dispersion relation}.---The eigenvalue equations for the Hamiltonian in Eq.~(\ref{array_Hamiltonian}) are $H_{\rm in}|\Psi_{{\rm R},l}\rangle=E_l|\Psi_{{\rm R},l}\rangle$ and $H^{\dagger}_{\rm in}|\Psi_{{\rm L},l}\rangle=E_l^*|\Psi_{{\rm L},l}\rangle$, where $|\Psi_{{\rm R},l}\rangle$ and $|\Psi_{{\rm L},l}\rangle$ are the right and left vectors of the $l$th eigenstate with $l=1,\cdots, 2N$ in the biorthogonal representation~\cite{JPhys.A47}. The localization of the eigenstates can be studied via the expectation value of the projection operator~\cite{PhysRevLett.121.026808} $\Pi_n=|A_n\rangle\langle A_n|+|B_n\rangle\langle B_n|$ onto the $n$th unit cell. We find~\cite{SupplementalMaterial} that the expectation values $\langle \Psi_{\rm L,left}|\Pi_n|\Psi_{\rm R,left} \rangle = \mathcal{N}_{\rm L}^*\mathcal{N}_{\rm R}(\xi_{\rm L} \xi_{\rm R})^{n}$ for the left edge state $|\Psi_{\rm L(R),left} \rangle$ and $\langle \Psi_{\rm L,right}|\Pi_n|\Psi_{\rm R,right} \rangle = \mathcal{N}_{\rm L}^*\mathcal{N}_{\rm R}(\xi_{\rm L} \xi_{\rm R})^{N+1-n}$ for the right edge state $|\Psi_{\rm L(R),rigth} \rangle$, where $|\Psi_{\rm L(R),rigth(left)} \rangle \in \{ |\Psi_{{\rm L(R)},l} \rangle \}$, $\mathcal{N}_{\rm L}$ and $\mathcal{N}_{\rm R}$ are normalization constants, coefficients $\xi_{\rm L}$ and $\xi_{\rm R}$ are related to the left and right eigenvectors~\cite{SupplementalMaterial}. The subscripts ``left" and ``right" mean the left and right edge states, respectively. The condition  $|\xi_{\rm L} \xi_{\rm R}|<1$ denotes the existence of exponentially localized edge states. The topological phase transition point corresponding to Eq.~(\ref{array_Hamiltonian}) is given via the condition $|\xi_{\rm L} \xi_{\rm R}|=1$ as~\cite{SupplementalMaterial}
\begin{equation}
\tilde{t}_A^{\;2}+\gamma_1^2=\tilde{t}_B^{\;2}+\gamma_2^2 , \label{transitonpoint}
\end{equation}
which is determined by $\varphi$ and $\gamma_{i}$ with $i=1,\,2$, and reduced to $\tilde{t}_A=\tilde{t}_B$ for $\gamma_{1}=\gamma_{2}=0$ corresponding to the Hermitian Hamiltonian. $\tilde{t}_A<\tilde{t}_B$ corresponds to topological phase and $\tilde{t}_A >\tilde{t}_B$ corresponds to topologically trivial phase when $\gamma_{1}=\gamma_{2}=0$. If the dissipative couplings are included, as shown in Fig.~\ref{figure2}(a), then the non-Hermitian Hamiltonian is in topological phase for $\tilde{t}_A^{\;2}+\gamma_1^2<\tilde{t}_B^{\;2}+\gamma_2^2$ and topologically trivial phase for $\tilde{t}_A^{\;2}+\gamma_1^2 > \tilde{t}_B^{\;2}+\gamma_2^2$. That is, the environment changes the phase diagram. For example, topologically trivial phase with $\tilde{t}_A > \tilde{t}_B$ in the Hermitian Hamiltonian may be changed to non-trivial one by the environment when $\gamma_{2} > \sqrt{\tilde{t}_A^{\;2}-\tilde{t}_B^{\;2}+\gamma_1^2}$.

To further observe the phase transition point and dispersion relation changed by the dissipative coupling, we transform the Hamiltonian $H_{\rm in}$ in Eq.~(\ref{array_Hamiltonian}) into the momentum space~\cite{SupplementalMaterial} $H_k=(t_A+t_B \cos k )\sigma_x + i t_B \sin k \sigma_y$, which satisfies the chiral symmetry $\sigma_z H_k \sigma_z=-H_k$.  Here, $\sigma_{x,y,z}$ are Pauli matrices. Thus, the dispersion relations are $E_{\pm}(k)=\pm\sqrt{t^2_{A}+ t^2_{B}+2t_{A}t_{B}\cos k}$. As shown in Fig.~\ref{figure2}(b), $E_{\pm}(k)$ return to the case of the Hermitian Hamiltonian when $\gamma_{1}=\gamma_{2}=0$, and two bands are connected at the phase transition point $\varphi=\pi/2$ for $k=\pm\pi$. However, when the dissipative couplings are included, the real parts of $E_{\pm}(k)$ are connected, but the imaginary parts of $E_{\pm}(k)$ are gapped (connected) at $\varphi=\pi/2$ with $k=\pm\pi$  in Fig.~\ref{figure2}(c) for $\gamma_{1}\neq\gamma_{2}$ (Fig.~\ref{figure2}(d) for $\gamma_{1}=\gamma_{2}$). The winding number~\cite{PhysRevA.97.052115,arXiv.1906.04700v1} $\nu_E=\int dk \partial_k \text{Arg}[E_{\pm}(k)]/{2 \pi}$, one of the topological invariants, is calculated as (see Ref.~\cite{SupplementalMaterial})
\begin{equation*}
\nu_E= \begin{cases}
0, & \tilde{t}_A^{\;2} +\gamma_{1}^2 \ge \tilde{t}_B^{\;2}+\gamma_{2}^2, \\
1, & \tilde{t}_A^{\;2}+\gamma_{1}^2 < \tilde{t}_B^{\;2}+\gamma_{2}^2,
\end{cases}
\end{equation*}
which agrees with the result given in Eq.~(\ref{transitonpoint}).

\begin{figure}
\includegraphics[scale=0.08]{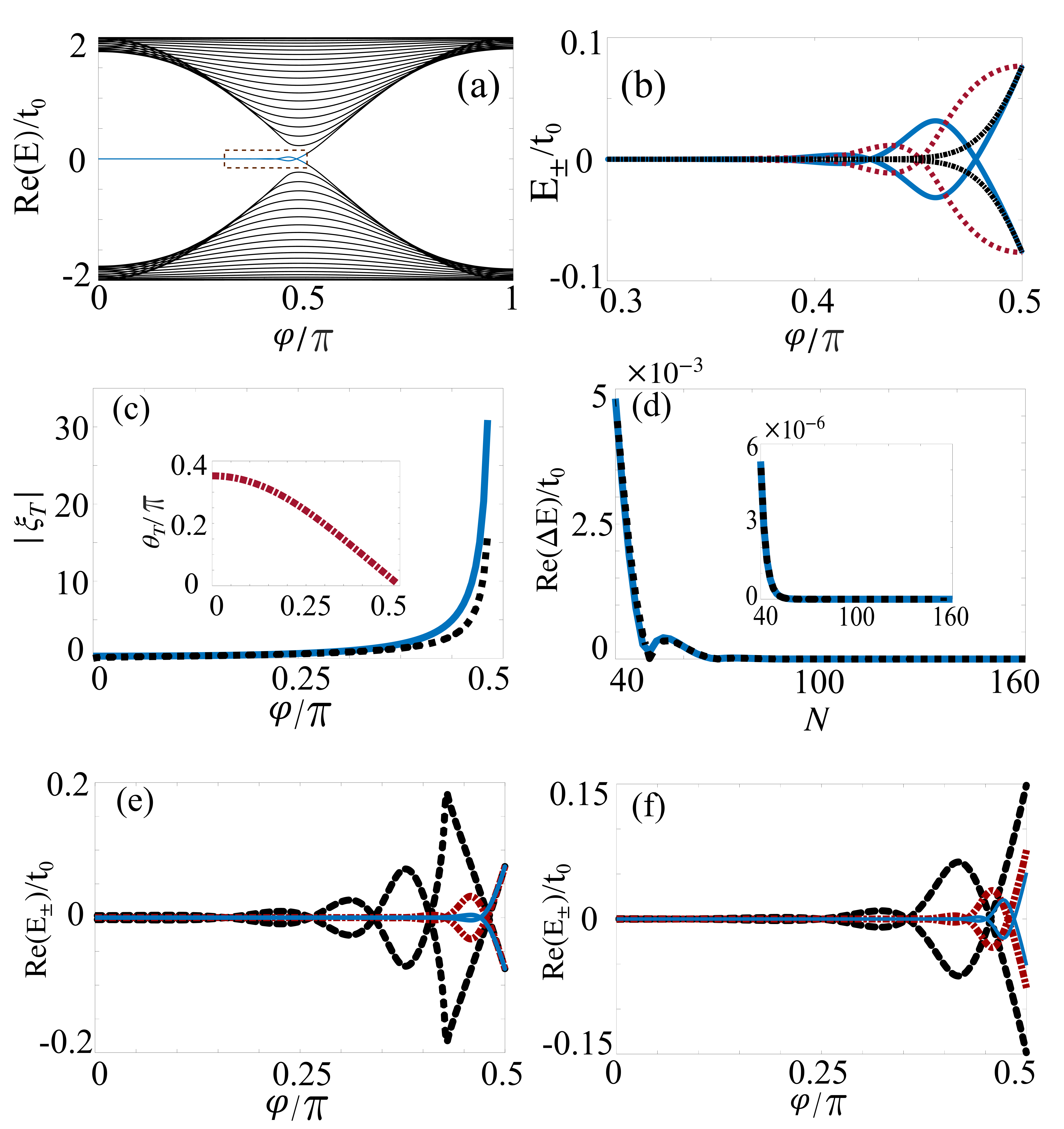}
\caption{(a) Real parts of the eigenvalues $E$ corresponding to the non-Hermitian Hamiltonian versus $\varphi$ for $\gamma_{1}/t_0=\gamma_{2}/t_0=1, N=20$. Real parts of the eigenvalues $E_{\pm}$ for hybridized edge states corresponding to blue solid curves in the dotted box are enlarged in (b),  red dashed curves in (b) correspond to the imaginary parts of $E_{\pm}$. Black chain-dotted curves correspond to values of $E_{\pm}$ with $\gamma_{1}=\gamma_{2}=0$. (c) Localization length ($1/\log|A_T|$) of $\xi_{T}$ for parameters $\gamma_{1}/t_0=\gamma_{2}/t_0=1$ (blue solid)  and parameters $\gamma_{1}=\gamma_{2}=0$ (black dotted) when $N=20$. $\theta_T$ is plotted as a function of $\varphi$ in inset figure. (d) Numerical (blue solid) and analytical (black dashed) results for ${\rm Re} (\Delta E)$ versus $N$ with $\varphi=0.4\pi, \gamma_{1}/t_0=\gamma_{2}/t_0=1$. Inset figure is ${\rm Re} (\Delta E)$ as a function of $N$ when $\varphi=0.4\pi$ and $\gamma_{1}=\gamma_{2}=0$, which is the result of the Hermitian SSH chain. (e) Numerical results of ${\rm Re}(E_{\pm})$ as a function of $\varphi$ for $\gamma_{1}/t_0=\gamma_{2}/t_0=0.5$ (blue solid), 1 (red chain-dotted), and 2 (black dashed), respectively, with $N=20$. (f) Numerical results of ${\rm Re}(E_{\pm})$ as a function of $\varphi$ with $\gamma_{1}/t_0=\gamma_{2}/t_0=1$ and $N=10$ (black dashed line), 20 (red chain-dotted line), and 30 (blue solid line), respectively.}\label{figure3}
\end{figure}

\textit{Edge states with Majorana-like oscillation}.---The real parts of the eigenvalues corresponding to Eq.~(\ref{array_Hamiltonian}) are plotted versus $\varphi$ in Fig.~\ref{figure3}(a). In contrast to those of the Hermitian Hamiltonian, we find from Figs.~\ref{figure3}(a) and (b) that both real and imaginary parts of the nonzero eigenvalues corresponding to hybridized edge states exhibit oscillation. This oscillation is very similar to the Majorana one~\cite{PhysRevLett.122.147701,Nature.531.206}, which may be used for transporting or identifying Majorana modes. Here, the oscillation demonstrates the nontrivial modification of edge states by dissipative couplings.

To analyze the relation between the Majorana-like oscillation and dissipative couplings, we write hybridized states for the right vectors of edge states~\cite{PhysRevLett.124.023603}
\begin{equation}
	|\Psi_{\rm R,\pm} \rangle= \frac{1}{\sqrt{2}}\left(|\Psi_{\rm R,left}\rangle \pm  |\Psi_{\rm R,right}\rangle\right).
\end{equation}
Similarly,  the hybridized states $\langle \Psi_{\rm L,\pm}|$ for the left vectors of edge states can be obtained. Thus, the eigenenergies for these states are $E_{\pm} =\langle\Psi_{\rm L,\pm}|H_{\rm in}|\Psi_{\rm R,\pm}\rangle=\pm\mathcal{N}_{\rm L}^*\mathcal{N}_{\rm R}\left(\tilde{t}_A-i\gamma_{1}\right)^{N+2}/\left(i\gamma_{2}-\tilde{t}_B\right)^{N+1}$,
which are simplified to~\cite{SupplementalMaterial,LectureNotesVol.919}
\begin{equation}
E_{\pm} = \zeta \exp \left(-\frac{N-1}{\xi_T}\right), \label{oscillation_eq}
\end{equation}
for the large number $N$, with the constant $\zeta$, determined by the normalization condition~\cite{SupplementalMaterial}. The parameter $\xi_T = 1/(\ln |A_T| + i\theta_T)$ characterizes the effective complex localization length of the non-Hermitian edge states with $A_T=\sqrt{(\tilde{t}_B^2+\gamma_2^2)/(\tilde{t}_A^2+\gamma_1^2)}$ and $\theta_T={\rm Arg} \left[\left(\tilde{t}_B -i\gamma_2\right)/\left(\tilde{t}_A - i\gamma_1\right)\right]$. Due to this feature, the energy variations of hybridized non-Hermitian edge states are different from those of the Hermitian ones. The energy splitting $\Delta E=E_{+}-E_{-}$ between two hybridized non-Hermitian edge states is
\begin{equation}
\Delta E  \propto [\cos{(N_T\theta_T)} - i\sin{(N_T\theta_T)}] e^{-N_T\ln|A_T|}, \label{Eq9}
\end{equation}
with $N_T=N-1$. The dissipative couplings nontrivially change the edge states such that the hybridized edge states show interesting properties. The coherent part ${\rm Re}\left(\Delta E\right) \propto \cos{\left(N_T\theta_T\right)} \exp\left(-N_T\ln |A_T|\right)$ has the similar expression as that for the Majorana oscillation described in Ref.~\cite{PhysRevB.86.220506}. Here, $N_{T}$, $1/\ln|A_T|$, and $\theta_T$ are equivalent to the length of the nanowire, the localization strength, and effective Fermi wave vector in Ref.~\cite{PhysRevB.86.220506}, respectively. Moreover, the topological-control parameter $\varphi$ is equivalent to the magnetic field for nanowires to tune the hybridization between Majorana modes.

In Fig.~\ref{figure3}(c), we show  that $\theta_T$ varies from $0.4\pi$ to $0$ when $\varphi$ changes from $0$ to $\pi/2$ in the topological phase. The parameter regime for $\theta_T$ can be larger for proper dissipative couplings. We also find that the complex localization length $\xi_{T}$ has the tendency as that of the Hermitian SSHc when $\varphi$ varies from $0$ to $\pi/2$. For the larger vaule of $\varphi$, the localization length becomes larger, yielding strong hybridization between two edge states. As shown in Fig.~\ref{figure3}(d), the analytical expression in Eq.~(\ref{oscillation_eq}) agrees well with the numerical result when $N$ is large.

The dissipative couplings change the Majorana-like oscillation by increasing amplitude and period of the oscillation as shown in Fig.~\ref{figure3}(e). For a given length of the SSHc in Eq.~(\ref{Eq9}), when $\varphi$ is taken as a value in the topologically nontrivial phase, larger values of $\gamma_1$ and $\gamma_2$ result in small value of $\log |A_T|$ when $\gamma_{1}=\gamma_{2}$, and thus the amplitude related term $\exp\left(-N\ln |A_T|\right)$ becomes larger. Also, large dissipative couplings result in less zero points of the oscillatory term $\cos{\left(N_T \theta_T \right)}$, and thus the oscillations also become less. On the other hand, for the given dissipation rates, as shown in Fig.~\ref{figure3}(f), the larger $N$ results in smaller oscillation amplitude. When the chain becomes very long, e.g., $N \ge 100$, the oscillation amplitude is negligibly small. That is, the edge states are topologically protected for a long chain even though there are dissipative couplings induced by the common environments.
We note that the phenomena shown above for $\gamma_{1}=\gamma_{2}$ are also valid for $\gamma_{1}\neq\gamma_{2}$.

\begin{figure}
\includegraphics[scale=0.28]{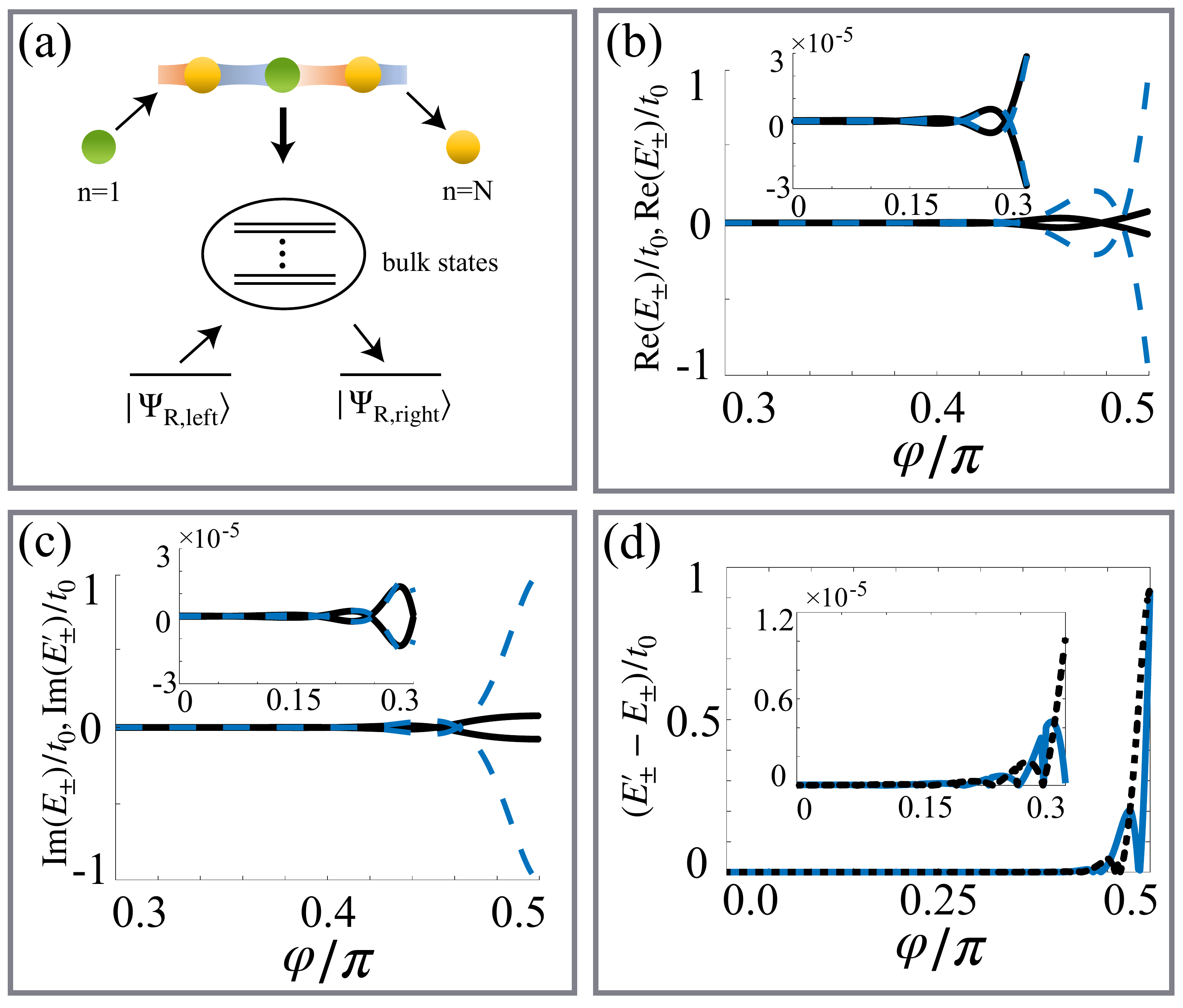}
\caption{(a) Schematics of the indirect coupling between the first and last qubits and the coupling between two edge states; (b) Real parts of $E_{\pm}^{\prime}$ (blue dashed curves) and $E_{\pm}$ (black solid curves) as a function of $\varphi$; (c) Imaginary parts of $E^{\prime}_{\pm}$ (blue dashed curves) and $E_{\pm}$ (black solid curves) as a function of $\varphi$; (c) $|{\rm Re}(E_{\pm}^{\prime}-E_{\pm})|$ (blue solid curves) and  $|{\rm Im}(E_{\pm}^{\prime}-E_{\pm})|$ (black dashed curves) as a function of $\varphi$. Here, we take $\gamma_{1}/t_{0}=\gamma_{2}/t_{0}=1$, $\tau/t_0=2$, $N=20$.}\label{figure4}
\end{figure}

\textit{Topology induced interaction between two edge qubits and measurements}.---As shown in Fig.~\ref{figure4}(a), the two edge qubits in the SSHc are indirectly coupled through $2N-2$ middle qubits and their environments. The relation about the hybridized states and the interaction between two edge qubits can be analyzed via the master equation
\begin{equation}
	\frac{d}{dt} \rho = -i[ H_T + H_{\rm 1,middle} + H_{\rm N,middle}, \rho] + \mathcal{L} \rho. \label{eq:6}
\end{equation}
Here, $H_T = H_a+H_b+H_{\rm middle}$ with $H_a = \omega \hat{a}_1^\dagger\hat{a}_1$, $H_b = \omega \hat{b}_N^\dagger\hat{b}_N$ and $H_{\rm middle} = \sum_{i=1}^{N-1}(\omega \hat{a}_{i+1}^\dagger\hat{a}_{i+1} + \omega \hat{b}_i^\dagger \hat{b}_i) + \sum_{i=1}^{N-1}(\tilde{t}_B \hat{a}_{i+1}^\dagger \hat{b}_i + {\rm H.c.}) + \sum_{i=2}^{N-1}(\tilde{t}_A \hat{a}_i^\dagger \hat{b}_i + {\rm H.c.})$. Moreover, $H_{\rm 1,middle} = \tilde{t}_A(\hat{a}_1^\dagger\hat{b}_1 + {\rm H.c.})$ and $H_{\rm N,middle} = \tilde{t}_A (\hat{a}_N^\dagger\hat{b}_N + {\rm H.c.})$ denote the coupling between the qubits A and B in first and last unit cells, respectively. After neglecting the effect of the independent environment on each qubit~\cite{SupplementalMaterial}, we write the Lindblad operator as $\mathcal{L}\rho = \sum_{i=1}^{N}\tau\mathcal{L}[\hat{a}_i+\hat{b}_i]\rho + \sum_{i=1}^{N-1}\tau\mathcal{L}[\hat{a}_{i+1}+\hat{b}_i]\rho$ for the same dissipative coupling strength with $\mathcal{L}[\hat{o}]\rho=\hat{o}\rho \hat{o}^\dagger - \frac12 \hat{o}^\dagger \hat{o}\rho -\frac12 \rho \hat{o}^\dagger \hat{o}$, and the coefficient $\tau$ refers to the dissipation rate of either qubit A or B through the waveguide (or cavity).

In the rotating reference frame with the frequency $\omega$, equation~(\ref{eq:6}) can be rewritten as
\begin{equation}
	\frac{d}{dt}\rho = (\mathcal{L}[\hat{a}_1] + \mathcal{L}[\hat{b}_N] + \mathcal{L}_{\rm middle} + J + \mathcal{L}_{\rm int})\rho, \label{motion_new}
\end{equation}
where the first two terms represent local dissipations for the first and last qubits. The third and forth terms read $\mathcal{L}_{\rm middle}\rho = -i(\vec{a^\dagger})^T M \vec{a}\rho + i\rho(\vec{a^\dagger})^T M^*\vec{a}$ and $J\rho = 2 \vec{a}^T M^{\prime} \rho \vec{a^\dagger}$, respectively. Here, we consider the vector $\vec{a}=(\hat{b}_1, \hat{a}_2, \hat{b}_2,\dots \hat{a}_N)$, and $(2N-2)\times (2N-2)$ matrices $M$ and $M^{\prime}$ with entries $M_{i,i}=-i\tau, M_{2i-1,2i}=M_{2i,2i-1}=\tilde{t}_B-i\tau/2, M_{2i,2i+1}=M_{2i+1,2i}=\tilde{t}_A-i\tau/2$, $M^{\prime}_{i,j}=\tau \delta_{i,j} + \tau \delta_{i,j+1}/2 + \tau \delta_{i+1,j}/2$. The last term $\mathcal{L}_{\rm int}\rho$ describes the coupling between edge qubits and middle qubits which includes vectors $\vec{G}_j=(\tilde{t}_A-i\tau/2) \delta_{1,j}$, $\vec{V}_j=(\tilde{t}_A-i\tau/2) \delta_{2N-2,j}$.

After eliminating the $2N-2$ qubits in the middle of the SSHc~\cite{PhysRevA.102.013714,SupplementalMaterial}, the coupling between the first and last qubits can be described via an effective Hamiltonian
\begin{equation}
H_{\rm edge}=\left(\Delta g-i\Delta \gamma\right)(\hat{a}_1^\dagger\hat{b}_N + {\rm H.c.}),
\end{equation}
which has the eigenvalues~\cite{SupplementalMaterial} $E^{\prime}_{\pm}=\pm \left(\Delta g-i\Delta \gamma\right)$ with $\Delta g={\rm Re}[\vec{G}^T M^{-1}\vec{V}]$ and $\Delta\gamma={\rm Im}[\vec{G}^T M^{-1}\vec{V}]$. As shown in Figs.~\ref{figure4}(b) and (c), the real and imaginary parts of $E^{\prime}_{\pm}$ and $E_{\pm}$ obtained numerically are comparatively plotted. We find that they coincide for the small value of $\varphi$. However, as shown in Fig.~\ref{figure4}(d), their difference becomes larger when $\varphi$ approaches the phase transition point $\pi/2$, where the adiabatic elimination is broken down. This shows that the oscillation can be partially measured via dynamics of two edge qubits for small value of $\varphi$. The full measurement on the oscillation can be realized via a cavity~\cite{SupplementalMaterial}.

\textit{Discussions and conclusions}.---We show that dissipative couplings induced by common environments in SSHc make edge states have complex localization lengths, and result in the Majorana-like oscillation for the hybridized edge states in the finite-size chain. The dissipative coupling can also nontrivially change topological phase. In practice, the on-site loss always exists. We find~\cite{SupplementalMaterial} that the on-site loss only shifts imaginary parts of the eigenvalues corresponding to the non-Hermitian Hamiltonian and does not affect the oscillatory behavior. However, for the non-Hermitian Hamitlonian with gain and loss as the on-site potentials, the oscillations can be suppressed when the gain and the loss are alternately exerted on the qubits $A$ and $B$ of the SSHc. If the loss (gain) and the gain (loss) are only exerted on the first and the last qubits, then the oscillations completely disappear.

Our study here can be realized by various systems, e.g., superconducting quantum circuits, in which the nearest neighbour qubit pairs are coupled to waveguides or bad cavities, acting as common environments. The qubits and their environments can reach superstrong and even ultrastrong coupling~\cite{NatureRevPhys}, thus the parameters used here are reachable. Moreover, the couplings between the superconducting qubits are well controlled, thus it is not difficult to make the system change from the topological to non-topological phase. We finally emphasize that the dissipative couplings provide a new way to manipulate the edge states and may have potential applications in topological quantum computing.

\textit{Acknowledgments}.---Y.X.L. is supported by the National Basic Research Program (973) of China under Grant No. 2017YFA0304304 and the Key R\&D Program of Guangdong province under Grant No. 2018B030326001.

\pagebreak
\clearpage

\end{document}